\documentclass[tightenlines,superscriptaddress,twocolumn,prl,notitlepage,nofootinbib,preprintnumbers]{revtex4-1}

\usepackage{graphicx}
\usepackage{amsmath}
\usepackage{amsfonts}
\usepackage{amssymb}
\usepackage{float}
\usepackage{hyperref}
\usepackage{multirow}
\usepackage{bbold}
\usepackage{color}
\usepackage{soul}
\usepackage{makecell}
\usepackage{minibox}
\usepackage{soul}
\usepackage{fancyhdr}

\begin{document}

\preprint{LA-UR-16-29320}
\title{Determining reactor fuel type from continuous antineutrino monitoring}

\author{Patrick Jaffke}
\email{corresponding author: pjaffke@vt.edu}
\affiliation{Center for Neutrino Physics, Virginia Tech, Blacksburg, VA, USA}

\author{Patrick Huber}
\affiliation{Center for Neutrino Physics, Virginia Tech, Blacksburg, VA, USA}

\date{\today}

\begin{abstract}
We investigate the ability of an antineutrino detector to determine
the fuel type of a reactor. A hypothetical $5\,\mathrm{t}$
antineutrino detector is placed $25\,\mathrm{m}$ from the core and
measures the spectral shape and rate of antineutrinos emitted by
fission fragments in the core for a number of $90\,\mathrm{d}$
periods. Our results indicate that four major fuel types can be
differentiated from the variation of fission fractions over the
irradiation time with a true positive probability of detection at $\sim95\%$.
In addition, we demonstrate that antineutrinos can identify
the burn-up at which weapons-grade mixed-oxide (MOX) fuel would be
reduced to reactor-grade MOX on average, providing assurance that plutonium
 disposition goals are met. In addition, we investigate
removal scenarios where plutonium is purposefully diverted from a
mixture of MOX and low-enriched uranium (LEU) fuel. Finally, we
discuss how our analysis is impacted by a spectral distortion around
$6\,\mathrm{MeV}$ observed in the antineutrino spectrum measured from
commercial power reactors.
\end{abstract}

\maketitle


The end of the Cold War after the collapse of the Soviet Union in 1991
left the United States and Russia with a large number of surplus
nuclear weapons~\cite{pmda}. Ultimately, the plutonium contained in
these surplussed nuclear weapons needs to be disposed of. There are
various techniques proposed for the disposal of weapons plutonium, see
for instance Ref.~\cite{Hippel:2015}, and one of these techniques is
based on so-called mixed oxide fuel (MOX), where a large part of the
fissile content in regular reactor fuel is replaced with the
to-be-disposed plutonium. This MOX fuel can be used in commercial
light-water reactors and would thus allow one to convert some of the
plutonium to usable energy. The remaining plutonium which is not
fissioned will undergo a major change of its isotopic composition
rendering it less attractive for the use in nuclear weapons. Moreover,
whatever is left will be embedded in highly radioactive spent reactor
fuel, making retrieval expensive and difficult. MOX fuel is
successfully employed in Europe, in particular in France a significant
number of power plants are using MOX fuel on an ongoing basis. In the
MOX approach to plutonium disposal, the primary quantitative measure
of reaching the disposition goal is given by burn-up: fuel which has
reached a certain burn-up threshold will both have a significantly
changed mix of plutonium isotopes as well as be sufficiently protected
by its own radiation field. In this paper we investigate how
continuous antineutrino monitoring can be used as a complementary
verification technique of both disposal goals by directly measuring
the burn-up and ratio of plutonium-239 to plutonium-241, which serves
as a proxy for the fraction of plutonium-240; plutonium-240 is not
fissile and thus does not have its own, direct neutrino signature. In
addition, we study the hypothetical scenario of the intentional
removal of plutonium.

The monitoring of nuclear reactors via antineutrino emission was first
postulated nearly 40 years ago by Borovoi and
Mikaelyan~\cite{Mikaelyan:1978}. This concept has seen a recent
resurgence as a safeguards or verification
technique~\cite{Bernstein:2001cz,Bernstein08,Christensen:2013eza,Heeger:2012tc,Oguri:2014gta,Boireau:2015dda},
where antineutrinos offer the unique advantages of independence of
operation declarations and the ability to recover from a loss of
continuity~\cite{PhysRevLett.113.042503}. This type of reactor
monitoring would require surface-level detector technology, which has
yet to be demonstrated with sufficient fidelity but is the current goal
of many short-baseline neutrino experiments~\cite{Bowden:2016ntq}.

Antineutrino monitoring relies on the fact that fissions of different
fissile nuclides, such as $^{235,238}$U or $^{239,241}$Pu, produce
different spectral shapes in antineutrino energy. An overall
measurement of the rate of antineutrinos will determine the power of
the reactor, while a spectral decomposition can infer the core
content. These techniques were employed
previously~\cite{Christensen:2013eza,PhysRevLett.113.042503} to study
the capabilities of antineutrinos based on real-world
scenarios. We use the same process in this work.

Typically, the fuel evolution of the reactor of interest is simulated,
and thus the fission rates throughout the irradiation cycle are
obtained.  This allows one to compute the total antineutrino spectrum by
weighting these fission rates with the appropriate antineutrino yields
from a single fission of each fissioning isotope. The uncertainties in these
yields can be reduced with a previous calibration of the antineutrino
detector to a core with known composition and we assume this
calibration has been performed. The total antineutrino signal
represents the `observed' spectrum in our simulated experiment. The
expected events are separated into energy bins to acquire the spectral
shape, represented by
\begin{equation}
n_i = N \int_{E_i-\Delta E/2}^{E_i+\Delta E/2} \sigma(E) \vec{\mathcal{F}}\cdot \vec{S}(E)\mathrm{d}E,
\label{eq:events}
\end{equation}
with the width of the energy bin $\Delta E$, the interaction
cross-section $\sigma(E)$~\cite{Vogel:1999zy}, the fission rate vector
$\vec{\mathcal{F}}$, and the vector of antineutrino yields from each
fissile $\vec{S}(E)$~\cite{Huber:2011wv,Mueller:2011nm}. The
normalization $N$ takes into account the detector size,
location\footnote{We assume a baseline short enough to avoid neutrino
  oscillations via active neutrinos.}, and overall efficiency. The
choice of $\Delta E$ must be small enough to allow for good resolution
in the spectral shape. We have chosen $\Delta E=250\,\mathrm{keV}$ and
a detection threshold of $2\,\mathrm{MeV}$. Previous large-scale
experiments have demonstrated the ability to reach this level of
energy resolution~\cite{An:2015nua,RENO:2015ksa,Crespo-Anadon:2014dea}
and future short-baseline detectors are aiming to match or exceed
this, see for instance Ref.~\cite{Ashenfelter:2015uxt}.

One can compute a log-likelihood ratio by comparing the observed
spectrum, created by weighting the simulated fission rate vector
$\vec{\mathcal{F}}_S$ with the fluxes $\vec{S}(E)$, to the expected
spectral shape of Eq.~\ref{eq:events}. Minimizing the resulting
$\chi^2$-function, given below
\begin{equation}
\chi^2(\vec{\mathcal{F}}) = \displaystyle\sum_{i}^N \frac{(n_i(\vec{\mathcal{F}}) - n_i^\prime)^2}{n_i^\prime},
\label{eq:chisqr}
\end{equation}
provides the best-fit fission rate vector $\vec{\mathcal{F}}$ (or
maximum likelihood estimate), where the observed events from
$\vec{\mathcal{F}}_S$ in bin $i$ are $n_i^\prime$. This measurement of
the fission rate vector is then used to determine the core type
and progression along its irradiation cycle. Detection
statistics are simulated by randomizing the $n_i^\prime$ with a Poisson
distribution. We assume detection statistics dominate the error budget as
precise detector calibration and simulation have been achieved below the
few percent level and accurate background measurements have been incorporated
into previous calibration techniques~\cite{An:2016ses}.

The simulated reactor is a Westinghouse-style light water reactor
(LWR) loaded with various core compositions. The details of the
simulation and the core configuration and characteristics are given in
Ref.~\cite{Erickson:2016sdm}.  Our analysis is primarily concerned
with four core types. The first is weapons-grade MOX (WGMOX) used in
the LWR, corresponding to the actual disposition case. The second is
reactor-grade MOX (RGMOX), which corresponds the plutonium vector of
discharged uranium-based fuel, and usually is part of a fuel cycle
which includes reprocessing of spent fuel. The third fuel type is a
mixture of two-thirds low-enriched uranium (LEU) fuel and one third of
WGMOX, where one third corresponds approximately to the MOX fraction
used in France\footnote{Obviously, France is using RGMOX.}. Finally,
the fourth core is a full LEU core. Initial fuel compositions are
described and provided in Ref.~\cite{Erickson:2016sdm} for the various
cores considered here.  All four cores are simulated to run for a
total of $500\,\mathrm{d}$ of irradiation at full power, corresponding
to a burn-up of
$21\,\mathrm{MW}\,\mathrm{d}/\mathrm{kg}\,\mathrm{HM}$. Thus, the
fission rates from these simulations have an implicit time-dependence
$\vec{\mathcal{F}}(t)$ and Eq.~\ref{eq:events} is modified with an
integration over the detection time $T$. This implies that the
observed spectral shape will depend on {\it when} the detector
monitors the reactor.

\begin{figure}[h]
\centering
\includegraphics[width=\columnwidth]{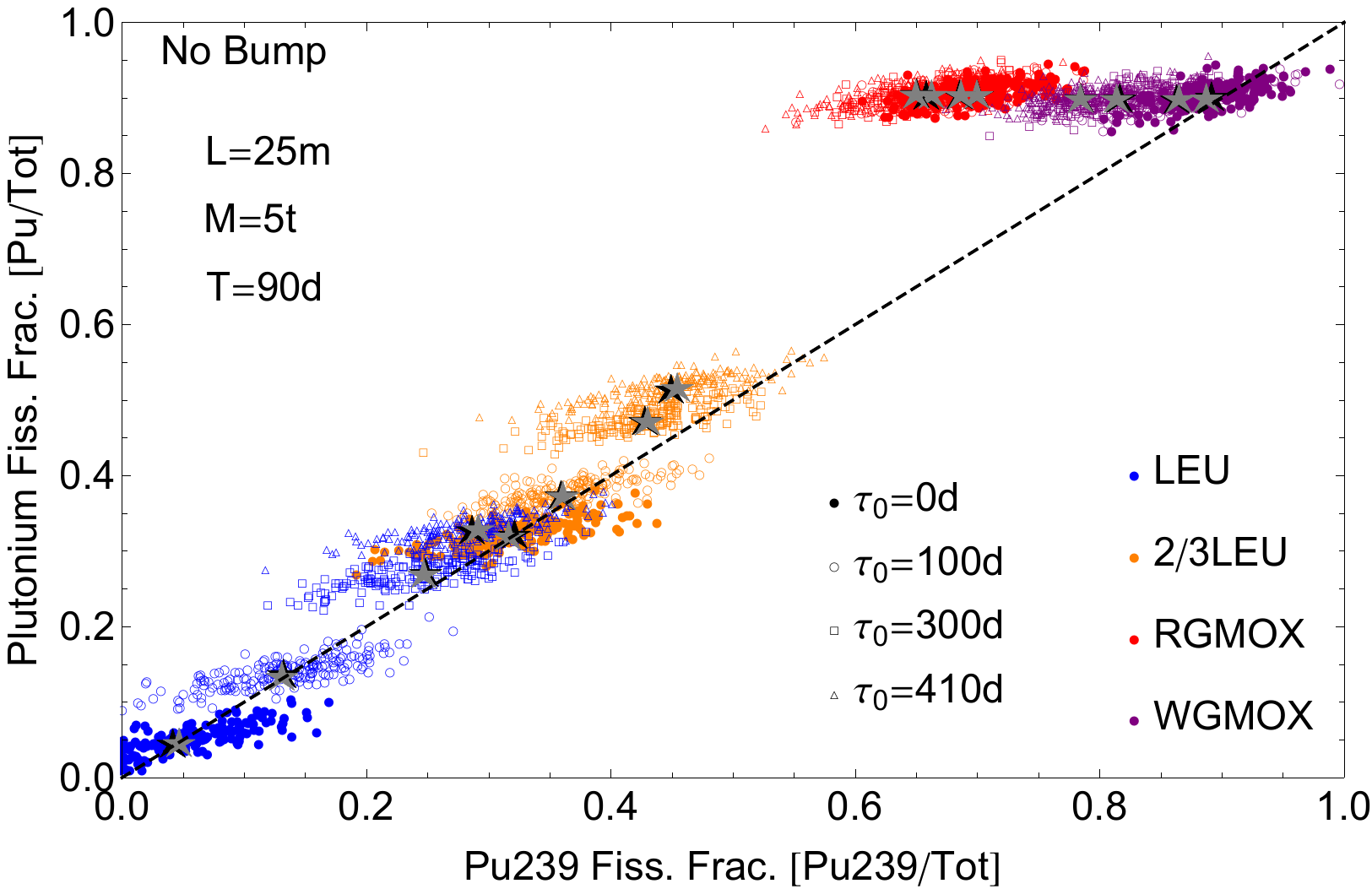}
\caption{\label{fig:Pu239vsPu} Plot of the derived best-fit plutonium
  and plutonium-239 fission fractions for four different fuel inventories
  and four different detection periods. The progression of fuel is
  illustrated by the movement of the centroids (stars) for
  each set. Black centroids are calculated from the data points and
  the gray are directly from the simulated fission rates. The dashed
  line is the physically-real plutonium boundary.}
\end{figure}

Two hundred cases were simulated for each of the four core types and
four different detection periods. These individual cases represent
different Poisson-randomized antineutrino spectra from the simulated
fission rates $\vec{\mathcal{F}}_S$. Each case was minimized via
Eq.~\ref{eq:chisqr} producing a best-fit core-averaged $\vec{\mathcal{F}}$ for that
detection period and fuel content. One can evaluate the best-fit
fission rate vector in various forms, such as the total plutonium
fission fraction versus the plutonium-239 fission fraction as in
Fig.~\ref{fig:Pu239vsPu}.
We can see that within the first $90\,\mathrm{d}$ measurement
(i.e. the start time is $\tau_0=0\,\mathrm{d}$) the cores appear as
expected. The pure LEU core (blue) contains a majority of uranium
fission and very little plutonium. The core with one third WGMOX
(orange) contains a sizable amount of plutonium initially and the
plutonium is a high-percentage plutonium-239, as the centroid falls near
the pure plutonium-239 dashed line. For the MOX cores, the reactor-grade
plutonium (red) begins far from the purity line and has mostly
plutonium fissions. The WGMOX core (purple) begins on the purity line
and also has a majority of plutonium fissions. The black centroids are
found by evenly weighting all data points in a particular set,
including those that fall in non-physical regimes, such as points beyond
the pure plutonium-239 line. The gray centroids mark the time-averaged
value of the fission quantity during its irradiation period, directly
calculated from the simulated time-dependent fission
rates.

From Fig.~\ref{fig:Pu239vsPu} we note that the initial state of the
mixed LEU+WGMOX core looks nearly identical to that of the last
detection period of the pure LEU core. This degeneracy can be broken
when we consider continuous antineutrino monitoring. With continuous
monitoring, one would know the starting irradiation time and relative
power from the rate measurement. With multiple measurements of the
antineutrino spectrum one could infer core inventory based on the
trajectory of two consecutive measurements, say $\tau_0=0\,\mathrm{d}$
and $\tau_0=100\,\mathrm{d}$. A trajectory that moves along the pure
plutonium-239 line is comprised of mostly LEU. Horizontal
shifts are cores that contain mostly MOX. We also note that the
antineutrino-derived fission fractions (black stars) are within a few percent
of the simulated values (gray stars) in Fig.~\ref{fig:Pu239vsPu}.

To acquire information on the plutonium grade, an important factor in
weapons-production~\cite{Carlson:1999}, one can use a ratio of the
plutonium fission rates. We use the ratio of plutonium-239 fission
rate to the total plutonium fission rate, which we label as the
plutonium fission grade ($G_\mathrm{Pu}$). Using $G_\mathrm{Pu}=90\%$
as the disposition goal\footnote{The thermal fission cross section of
  $^{241}$Pu is 35\% higher than the one of $^{239}$Pu, thus in a
  mixture of 93\% $^{239}$Pu and 7\% $^{241}$Pu 90\% of all fission
  will take place in $^{241}$Pu. }, we find based on the antineutrino
measurement alone that $16\%$ ($31\%$) of WGMOX (LEU+MOX) cores
remained above this goal at the final ($410-500\,\mathrm{d}$)
measurement period. The sensitivity to downgrading the plutonium is
worse in the mixed core as the uranium absorbs some of the total
fission rates, thus slowing the progression of plutonium fissions from
weapons-grade to reactor-grade. In addition, this measurement
represents a core-averaged plutonium ratio, so a lower $G_\mathrm{Pu}$
requirement would be necessary to ensure all assemblies fall below the
weapons-barrier or an assembly-by-assembly technique would need to be
employed as well~\cite{abdurrahman1993spent}.

The trajectory and the absolute distance traveled in the fission
quantity plane can be calculated from Fig.~\ref{fig:Pu239vsPu} and
compared between different cores and different measurement periods.
For example, the LEU cores both travel along the same trajectory, but
the pure LEU core has a much larger difference between two adjacent
measurement periods than the mixed core. This quantity, which we label
the differential burn-up, is given by
\begin{equation}
\Delta\vec{F} = \frac{\vec{F}(t_f) - \vec{F}(t_i)}{\vec{F}(t_f)},
\label{eq:DBA}
\end{equation}
where the fission fraction $\vec{F}(t)$ is created from the best-fit
values of the fission rate vector $\vec{\mathcal{F}}(t)$. We have
computed the differential burn-up between the four measurement periods
for each of the four simulated cores in Tab.~\ref{tab:DBA}.
\begin{table}[h]
\centering
\renewcommand{\arraystretch}{1.25}
    \begin{tabular}{|c||c|c|c|}
        \hline 
	\multicolumn{4}{|c|}{Differential Burn-up $\lbrack \Delta F_{\mathrm{Pu239}}$,$\Delta F_{\mathrm{Pu}} \rbrack$ in percent } \\ \hline \hline
	& $\tau_0=100\,\mathrm{d}$ & $\tau_0=300\,\mathrm{d}$ & $\tau_0=410\,\mathrm{d}$ \\ \hline
	LEU & $\lbrack218,201\rbrack$ & $\lbrack498,493\rbrack$ & $\lbrack589,616\rbrack$ \\ \hline
	$2/3$ LEU & $\lbrack11.8,16.1\rbrack$ & $\lbrack33.2,46.6\rbrack$ & $\lbrack39.9,59.5\rbrack$ \\ \hline
	RGMOX & $\lbrack-2.05,-0.01\rbrack$ & $\lbrack-5.96,-0.10\rbrack$ & $\lbrack-7.10,0.17\rbrack$ \\ \hline
	WGMOX & $\lbrack-3.20,-0.39\rbrack$ & $\lbrack-8.41,-0.25\rbrack$ & $\lbrack-12.0,-0.23\rbrack$ \\ \hline \hline
	Nominal & $\lbrack8.40,14.0\rbrack$ & $\lbrack30.0,38.0\rbrack$ & $\lbrack31.1,48.9\rbrack$ \\ \hline
	8 Rem. & $\lbrack10.1,14.6\rbrack$ & $\lbrack30.3,40.9\rbrack$ & $\lbrack40.9,54.2\rbrack$ \\ \hline
	20 Rem. & $\lbrack20.5,20.5\rbrack$ & $\lbrack44.9,52.1\rbrack$ & $\lbrack57.4,68.2\rbrack$ \\ \hline
    \end{tabular}
\caption{\label{tab:DBA} Differential burn-up analysis (DBA) of the
  plutonium-239 fission fraction $\Delta F_{\mathrm{Pu239}}$ and the
  plutonium fission fraction $\Delta F_{\mathrm{Pu}}$ between the
  initial $\tau_0=0\,\mathrm{d}$ measurement period and the three
  following measurement periods in percent. The DBA is conducted for
  the four simulated cores and is given as ordered pairs. Also listed
  are the DBA results for the removal cases where no fresh WGMOX assemblies
  (Nominal), $8$, or all $20$ are replaced with LEU fuel.}
\end{table}
The trends noted above appear quantitatively in this differential burn-up analysis (DBA). For
example, both LEU-type cores have a positive $\Delta F_{\mathrm{Pu}}$
across all time steps, indicating that these cores are producing
plutonium. However, the LEU core mixed with WG plutonium has a
significantly smaller $\Delta F_{\mathrm{Pu}}$ indicating that the
{\it rate} of plutonium production in the mixed core is dramatically
slower by about an order of magnitude for all time steps. In addition,
the grade of the plutonium decreases faster for the WGMOX core than
the RGMOX core. This difference is much more subtle and develops
slowly, but results in an almost doubling of $|\Delta
F_{\mathrm{Pu239}}|$, as can be seen in Tab.~\ref{tab:DBA}.

\begin{figure}[h]
\centering
\includegraphics[width=\columnwidth]{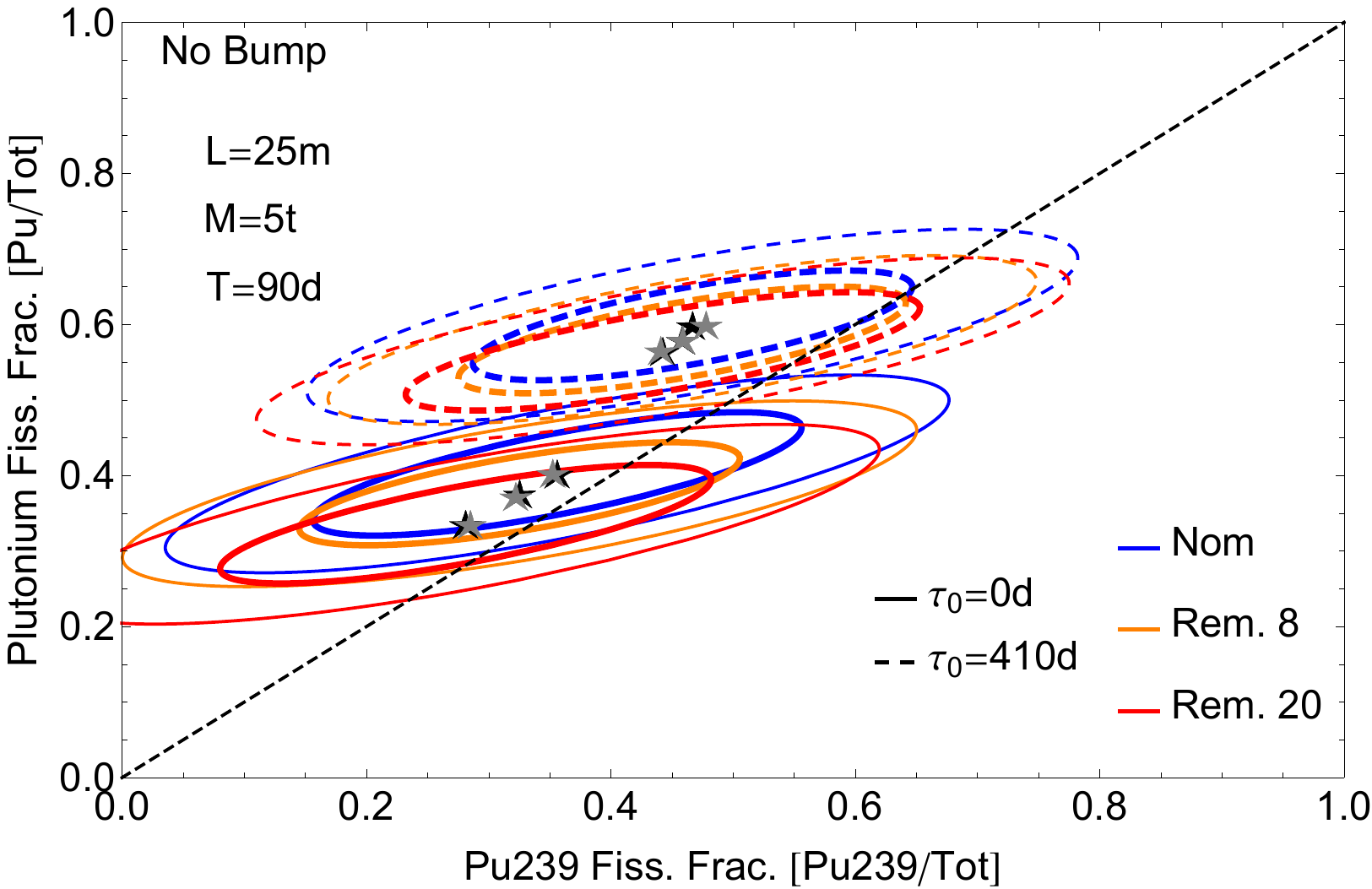}
\caption{\label{fig:Removal} Plot of the allowed region contours of the
  best-fit plutonium-239 and total plutonium fission fractions. The
  contours are derived by fitting Gaussian ellipses to the case
  distribution. The thick inner ellipse denotes the $1\sigma$ quartile
  and the thin outer ellipse the $2\sigma$ quartile. Centroids derived
  from the ellipses (black) and from the simulated fission rates
  (gray) are shown. The dashed line is the physically-real
  plutonium boundary.}
\end{figure}

Another possible scenario is the intentional removal of plutonium from
a core, such as the mixed LEU and WGMOX
core~\cite{Erickson:2016sdm}. The mixed core uses a total of $48$ MOX
assemblies, $28$ of which are once or twice-irradiated and therefore
not considered WGMOX any longer. This staggered burning is used in
reactor operation to flatten the neutron spectrum distribution.  We
investigate three scenarios: the first is the nominal run with no
removal of the fresh WGMOX assemblies. The second considers removing
$8$ fresh WGMOX assemblies from the periphery of the reactor and
replacing them with LEU assemblies. The fission rate at the edge of
the reactor core is relatively low and thus this case will be a
particular challenge for antineutrino monitoring. The final case
considers a full removal and replacement of all ($20$) fresh WGMOX
assemblies. We plot these scenarios and their exclusion contours in
the $F_\mathrm{Pu239}$~--~$F_\mathrm{Pu}$ plane for the initial and
final time-step in Fig.~\ref{fig:Removal}.
For the removal cases, we see that the differential burn-up vectors are
all aligned in relatively the same direction, but the magnitudes are
slightly different. One can also note that the differences between the
removal cases are much less pronounced than those for the full cores,
as seen in Tab.~\ref{tab:DBA}. Furthermore, it becomes apparent that
the equilibrium fission rates for each scenario are nearly identical,
making this difference more difficult to detect at later
irradiation times.

The analyses in Figs.~\ref{fig:Pu239vsPu},~\ref{fig:Removal}, and Tab.~\ref{tab:DBA}
demonstrate the process by which an antineutrino detector would infer the core content
of a nuclear reactor. Several time steps are needed to determine the
progression of the fuel and identify the differential burn-up. The sign
and magnitude resulting from DBA can distinguish between LEU, mixed
LEU and MOX, and MOX cores easily, but the difference between RGMOX
and WGMOX is more subtle. An absolute measurement of the fission
fractions can help to distinguish the latter cases. Next, we present
the sensitivity analysis for these absolute measurements.

\begin{figure}[h]
\centering
\includegraphics[width=\columnwidth]{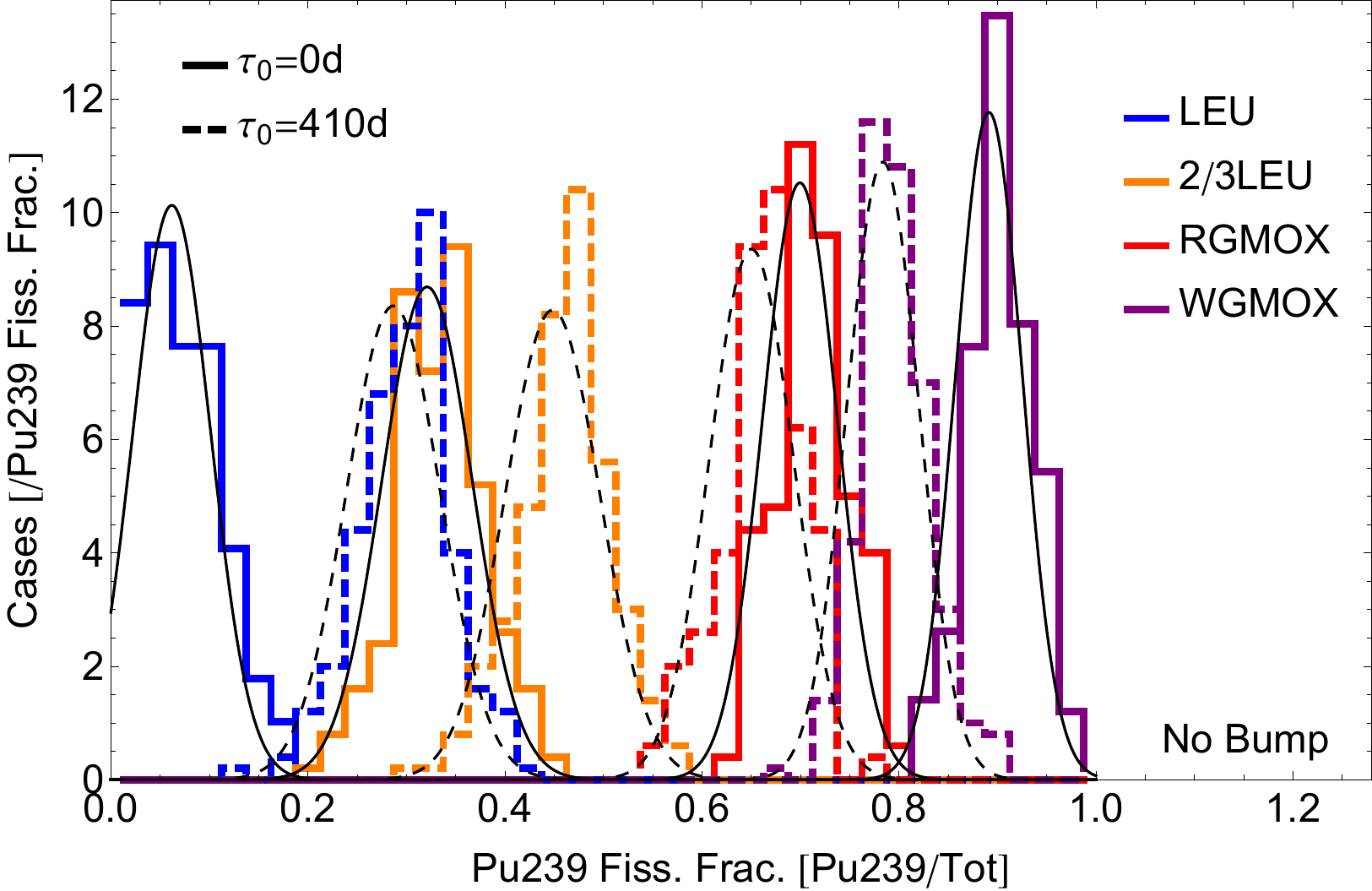}
\caption{\label{fig:Pu239FFHist} Histogram of the determined
  plutonium-239 fission fraction of the simulated cases for the first and
  last measurement times and four core compositions and the corresponding
  Gaussian distributions. This represents
  Fig.~\ref{fig:Pu239vsPu} projected onto its x-axis.}
\end{figure}

It is now possible to determine the true and false positive rates
(FPR) from the data generated so far. In Fig.~\ref{fig:Pu239FFHist},
we project the data onto the axis represented by the fission fraction
of plutonium-239. For each of the histograms we can determine
the parameters of a normal distribution, that is the mean and standard
deviation.  The resulting normal distributions are used to compute the
true and false positive rates as a function of the plutonium-239 fission fraction.
To allow a simple summary, we chose the critical value in
this variable such that the false negative and false positive rates are
equal and we will quote this common value. In the language of a receiver
operating characteristics, this corresponds to the balance point.

One can determine the false positive rate and the balance point for the four core and measurement
period distributions along the various fission fraction axes. First, the
simulated experiments are binned along an axis; Fig.~\ref{fig:Pu239FFHist}
provides the projection of Fig.~\ref{fig:Pu239vsPu} onto its x-axis of
plutonium-239 fission fraction.
As one would expect from Fig.~\ref{fig:Pu239FFHist}, we observe a very
small FPR during the first measurement period ($\sim0.1\%$ between the
LEU cores and $\sim0.4\%$ between the MOX cores). The second
measurement shows more significant FPR of $\sim1\%$ between both
the LEU cores and both the MOX cores. The third measurement begins to
show a higher FPR between the RGMOX and mixed LEU+WGMOX cores,
but only around $0.3\%$. The FPR between the LEU cores is $\sim3\%$
and between the MOX cores is $\sim1.8\%$. The last measurement shows
a FPR between the LEU cores at $4.6\%$, between the MOX cores at
$4.6\%$, and between the LEU+WGMOX and RGMOX cores at 
$\sim1.3\%$. The information for the first and last measurements is given
quantitatively in Tab.~\ref{tab:CoreSeparation}.

The low FPR indicate that only $\sim5\%$ of cases will be misidentified.
Multiple tests with different $\vec{F}(t)$ can isolate and enhance this
accuracy. For example, the plutonium or uranium fission fraction has very
good separation between the LEU and pure MOX cores, but is completely
unable to distinguish between RGMOX and WGMOX. Using the plutonium-239
grade, the FPR drops by about a factor of $3$ between the RGMOX
and WGMOX cores, but increases drastically for the LEU cores due to the
lower plutonium fission fractions. For the removal scenarios, the FPRs are
consistently above $50\%$ among all measurements, removal
scenarios, and fission fraction tests, implying that detection of those is
beyond the capabilities of antineutrino monitoring. This is primarily because
antineutrinos measure core-averaged quantities instead of individual
assemblies.

\begin{table*}[t]
\centering
\renewcommand{\arraystretch}{1.25}
    \begin{tabular}{|c||c||c|c|c||c|c|c||c|c|c|}
        \hline 
	& \multicolumn{10}{c|}{False Positive Rates} \\ \hline \hline
	& & \multicolumn{3}{c||}{No Bump} & \multicolumn{3}{c||}{U235 Bump} & \multicolumn{3}{c|}{Pu239 Bump} \\ \hline \hline
	& & LEU+MOX & RGMOX & WGMOX & LEU+MOX & RGMOX & WGMOX & LEU+MOX & RGMOX & WGMOX \\ \hline
	\multirow{3}{*}{First} & LEU & $0.119\%$ & $<10^{-4}\%$ & $<10^{-4}\%$ & $0.00380\%$ & $<10^{-4}\%$ & $<10^{-4}\%$ & $<10^{-4}\%$ & $<10^{-4}\%$ & $<10^{-4}\%$ \\ \cline{2-11}
	& LEU+MOX & & $<10^{-4}\%$ & $<10^{-4}\%$ & & $<10^{-4}\%$ & $<10^{-4}\%$ & & $<10^{-4}\%$ & $<10^{-4}\%$ \\ \cline{2-11}
	& RGMOX & & & $0.377\%$ & & & $0.0318\%$ & & & $<10^{-4}\%$ \\ \hline \hline
	\multirow{3}{*}{Last} & LEU & $4.60\%$ & $0.00290\%$ & $<10^{-4}\%$ & $0.612\%$ & $<10^{-4}\%$ & $<10^{-4}\%$ & $<10^{-4}\%$ & $<10^{-4}\%$ & $<10^{-4}\%$ \\ \cline{2-11}
	& LEU+MOX & & $1.31\%$ & $0.00385\%$ & & $0.0533\%$ & $<10^{-4}\%$ & & $<10^{-4}\%$ & $<10^{-4}\%$ \\ \cline{2-11}
	& RGMOX & & & $4.61\%$ & & & $1.21\%$ & & & $<10^{-4}\%$ \\ \hline
    \end{tabular}
\caption{\label{tab:CoreSeparation} Upper triangular portion of the
  core separation FPR matrices using the plutonium-239 fission fraction. Each
  upper triangular cell corresponds to the percentage of falsely
  identified cores between the various core compositions. This is
  done for the three considered locations of the spectral structure
  (no bump, a uranium-235 bump, and a plutonium-239 bump) and
  for the first ($\tau_0=0\,\mathrm{d}$) and last
  ($\tau_0=410\,\mathrm{d}$) measurement periods. The total FPR
  matrix is symmetric, as expected. A reduction in the FPR by about a
  factor of $3$ occurs between the RGMOX and WGMOX cores when using
  the plutonium grade as a diagnostic.}
\end{table*}

Recently, a spectral distortion was observed near $5\,\mathrm{MeV}$
(positron energy) in the reactor antineutrino flux of multiple
large-statistics
experiments~\cite{An:2015nua,RENO:2015ksa,Crespo-Anadon:2014dea}.
The origin of this so-called bump is unknown, but several theories have been
explored~\cite{Hayes:2015yka,Huber:2016fkt,Hayes:2016qnu}. Irrespective
of the bump explanation, one can verify the impact of the
spectral feature on the fuel determination by artificially placing the
observed bump in the converted antineutrino fluxes.

We choose to place the spectral feature in either uranium-235 or
plutonium-239. Using Fig.~2 in Ref.~\cite{An:2015nua}, the bump is
modeled as a Gaussian distribution where the mean and variance are
determined by a least-squares fit. We note that the normalization of
the Gaussian is fixed by the fission fractions and overall neutrino
rate in the relevant bins. The three Gaussian best-fit parameters vary
by only a small amount when fixing them instead by
Ref.~\cite{RENO:2015ksa} or Ref.~\cite{Crespo-Anadon:2014dea}. With
these new artificial fluxes, we repeat the process outlined above and
generate more sample simulations. As expected, a new spectral
distortion in either uranium-235 or plutonium-239 enhances the abilities of
an antineutrino detector to distinguish between fuel types. A bump in
uranium-235 has the effect of pinching the spread of $F_\mathrm{Pu}$ in
Fig.~\ref{fig:Pu239vsPu} as the uranium-235 fission rate, now easily
determined by the spectral distortion, essentially fixes the total
plutonium fissions. A bump in plutonium-239 pinches the distributions
along an axis in the $F_\mathrm{Pu239}$~--~$F_\mathrm{Pu}$ plane,
according to the core content. This also has the effect of reducing
the variance in the Gaussian distributions found in
Fig.~\ref{fig:Pu239FFHist}, which lowers the FPRs.
Overall, Tab.~\ref{tab:CoreSeparation} shows that this procedure
reduces the FPR by at least a factor of $3$ if the
uranium-235 spectrum contains the bump and below $10^{-4}\%$
for a plutonium-239 bump. The FPR rates remain at or above
$\sim 50\%$ for all removal scenarios across all measurement periods.
The sensitivity to downgrading is also enhanced by the bump location.
Less than $1\%$ ($20\%$) of the WGMOX (LEU+MOX) cores return
plutonium fission grades above weapons at the last measurement period
with a plutonium-239 bump.

This work has explored the abilities of a surface-deployed
antineutrino detector to determine the core composition of the reactor
it is monitoring via continuous spectral and rate measurements. The
spectrum, with an interaction threshold of $2\,\mathrm{MeV}$ and
binned into $250\,\mathrm{keV}$ bins, is fitted to an event
distribution providing best-fit values for the fission rates of
$^{235,238}$U and $^{239,241}$Pu. These best-fit fission rates are
combined into various fractions and we determine that for
$500\,\mathrm{d}$ of irradiation in a LWR, which will downgrade a full
WGMOX core to RGMOX, we can establish this average downgrade with $84\%$ confidence
based on antineutrino monitoring. For a one-third WGMOX two-thirds LEU
core we can determine the average downgrade with $69\%$ confidence. Multiple
measurement periods of $90\,\mathrm{d}$ within the irradiation cycle
can differentiate between our four major core compositions with $95\%$ accuracy. This is done
mostly by comparing the plutonium-239 fission fraction, but can be
reinforced with other fission fractions as well. Detecting the removal of
plutonium from a mixed LEU and WGMOX core with an antineutrino
detector is found to be incredibly difficult and highly reliant on the
first measurement period. The existence of a spectral distortion in either
the uranium-235 or plutonium-239 antineutrino fluxes only enhances the
monitoring capabilities mentioned above, except for the detection of
an intentional removal of plutonium.


\acknowledgements We would like to acknowledge A.~Bernstein, N.~Bowden
and A.~Erickson for useful discussions and comments on an earlier
version of this manuscript. This work was supported by the
U.S. Department of Energy National Nuclear Security Administration
under award \protect{DE-AC52-07NA27344} via sub-award
\protect{B612358} from Lawrence Livermore National Laboratory.

\bibliographystyle{apsrev} \bibliography{PMDA.bib}

\begin{thebibliography}{25}
\expandafter\ifx\csname natexlab\endcsname\relax\def\natexlab#1{#1}\fi
\expandafter\ifx\csname bibnamefont\endcsname\relax
  \def\bibnamefont#1{#1}\fi
\expandafter\ifx\csname bibfnamefont\endcsname\relax
  \def\bibfnamefont#1{#1}\fi
\expandafter\ifx\csname citenamefont\endcsname\relax
  \def\citenamefont#1{#1}\fi
\expandafter\ifx\csname url\endcsname\relax
  \def\url#1{\texttt{#1}}\fi
\expandafter\ifx\csname urlprefix\endcsname\relax\def\urlprefix{URL }\fi
\providecommand{\bibinfo}[2]{#2}
\providecommand{\eprint}[2][]{\url{#2}}

\bibitem[{\citenamefont{\protect{National Academy of Sciences}}(1994)}]{pmda}
\bibinfo{author}{\bibnamefont{\protect{National Academy of Sciences}}},
  \emph{\bibinfo{title}{{Management and Disposition of Excess Weapons
  Plutonium}}} (\bibinfo{publisher}{The National Academies Press},
  \bibinfo{year}{1994}).

\bibitem[{\citenamefont{von Hippel and MacKerron}(2015)}]{Hippel:2015}
\bibinfo{author}{\bibfnamefont{F.}~\bibnamefont{von Hippel}} \bibnamefont{and}
  \bibinfo{author}{\bibfnamefont{G.}~\bibnamefont{MacKerron}}
  (\bibinfo{year}{2015}).

\bibitem[{\citenamefont{Borovoi and Mikaelyan}(1978)}]{Mikaelyan:1978}
\bibinfo{author}{\bibfnamefont{A.}~\bibnamefont{Borovoi}} \bibnamefont{and}
  \bibinfo{author}{\bibfnamefont{L.}~\bibnamefont{Mikaelyan}},
  \bibinfo{journal}{Soviet Atomic Energy} \textbf{\bibinfo{volume}{44}},
  \bibinfo{pages}{589} (\bibinfo{year}{1978}), ISSN \bibinfo{issn}{0038-531X},
  \urlprefix\url{http://dx.doi.org/10.1007/BF01117861}.

\bibitem[{\citenamefont{Bernstein et~al.}(2002)\citenamefont{Bernstein, Wang,
  Gratta, and West}}]{Bernstein:2001cz}
\bibinfo{author}{\bibfnamefont{A.}~\bibnamefont{Bernstein}},
  \bibinfo{author}{\bibfnamefont{Y.-f.} \bibnamefont{Wang}},
  \bibinfo{author}{\bibfnamefont{G.}~\bibnamefont{Gratta}}, \bibnamefont{and}
  \bibinfo{author}{\bibfnamefont{T.}~\bibnamefont{West}}, \bibinfo{journal}{J.
  Appl. Phys.} \textbf{\bibinfo{volume}{91}}, \bibinfo{pages}{4672}
  (\bibinfo{year}{2002}), \eprint{nucl-ex/0108001}.

\bibitem[{\citenamefont{Bernstein et~al.}(2008)\citenamefont{Bernstein, Bowden,
  Misner, and Palmer}}]{Bernstein08}
\bibinfo{author}{\bibfnamefont{A.}~\bibnamefont{Bernstein}},
  \bibinfo{author}{\bibfnamefont{N.~S.} \bibnamefont{Bowden}},
  \bibinfo{author}{\bibfnamefont{A.}~\bibnamefont{Misner}}, \bibnamefont{and}
  \bibinfo{author}{\bibfnamefont{T.}~\bibnamefont{Palmer}},
  \bibinfo{journal}{J. Appl. Phys.} \textbf{\bibinfo{volume}{103}},
  \bibinfo{eid}{074905} (\bibinfo{year}{2008}),
  \urlprefix\url{http://scitation.aip.org/content/aip/journal/jap/103/7/10.1063/1.2899178}.

\bibitem[{\citenamefont{Christensen et~al.}(2015)\citenamefont{Christensen,
  Huber, and Jaffke}}]{Christensen:2013eza}
\bibinfo{author}{\bibfnamefont{E.}~\bibnamefont{Christensen}},
  \bibinfo{author}{\bibfnamefont{P.}~\bibnamefont{Huber}}, \bibnamefont{and}
  \bibinfo{author}{\bibfnamefont{P.}~\bibnamefont{Jaffke}},
  \bibinfo{journal}{Science and Global Security} \textbf{\bibinfo{volume}{23}},
  \bibinfo{pages}{20} (\bibinfo{year}{2015}), \eprint{1312.1959}.

\bibitem[{\citenamefont{Heeger et~al.}(2013)\citenamefont{Heeger, Littlejohn,
  Mumm, and Tobin}}]{Heeger:2012tc}
\bibinfo{author}{\bibfnamefont{K.~M.} \bibnamefont{Heeger}},
  \bibinfo{author}{\bibfnamefont{B.~R.} \bibnamefont{Littlejohn}},
  \bibinfo{author}{\bibfnamefont{H.~P.} \bibnamefont{Mumm}}, \bibnamefont{and}
  \bibinfo{author}{\bibfnamefont{M.~N.} \bibnamefont{Tobin}},
  \bibinfo{journal}{Phys. Rev.} \textbf{\bibinfo{volume}{D87}},
  \bibinfo{pages}{073008} (\bibinfo{year}{2013}), \eprint{1212.2182}.

\bibitem[{\citenamefont{Oguri et~al.}(2014)\citenamefont{Oguri, Kuroda, Kato,
  Nakata, Inoue, Ito, and Minowa}}]{Oguri:2014gta}
\bibinfo{author}{\bibfnamefont{S.}~\bibnamefont{Oguri}},
  \bibinfo{author}{\bibfnamefont{Y.}~\bibnamefont{Kuroda}},
  \bibinfo{author}{\bibfnamefont{Y.}~\bibnamefont{Kato}},
  \bibinfo{author}{\bibfnamefont{R.}~\bibnamefont{Nakata}},
  \bibinfo{author}{\bibfnamefont{Y.}~\bibnamefont{Inoue}},
  \bibinfo{author}{\bibfnamefont{C.}~\bibnamefont{Ito}}, \bibnamefont{and}
  \bibinfo{author}{\bibfnamefont{M.}~\bibnamefont{Minowa}},
  \bibinfo{journal}{Nucl. Instrum. Meth.} \textbf{\bibinfo{volume}{A757}},
  \bibinfo{pages}{33} (\bibinfo{year}{2014}), \eprint{1404.7309}.

\bibitem[{\citenamefont{Boireau et~al.}(2016)}]{Boireau:2015dda}
\bibinfo{author}{\bibfnamefont{G.}~\bibnamefont{Boireau}} \bibnamefont{et~al.}
  (\bibinfo{collaboration}{NUCIFER}), \bibinfo{journal}{Phys. Rev.}
  \textbf{\bibinfo{volume}{D93}}, \bibinfo{pages}{112006}
  (\bibinfo{year}{2016}), \eprint{1509.05610}.

\bibitem[{\citenamefont{Christensen et~al.}(2014)\citenamefont{Christensen,
  Huber, Jaffke, and Shea}}]{PhysRevLett.113.042503}
\bibinfo{author}{\bibfnamefont{E.}~\bibnamefont{Christensen}},
  \bibinfo{author}{\bibfnamefont{P.}~\bibnamefont{Huber}},
  \bibinfo{author}{\bibfnamefont{P.}~\bibnamefont{Jaffke}}, \bibnamefont{and}
  \bibinfo{author}{\bibfnamefont{T.~E.} \bibnamefont{Shea}},
  \bibinfo{journal}{Phys. Rev. Lett.} \textbf{\bibinfo{volume}{113}},
  \bibinfo{pages}{042503} (\bibinfo{year}{2014}).

\bibitem[{\citenamefont{Bowden et~al.}(2016)\citenamefont{Bowden, Heeger,
  Huber, Mariani, and Vogelaar}}]{Bowden:2016ntq}
\bibinfo{author}{\bibfnamefont{N.~S.} \bibnamefont{Bowden}},
  \bibinfo{author}{\bibfnamefont{K.~M.} \bibnamefont{Heeger}},
  \bibinfo{author}{\bibfnamefont{P.}~\bibnamefont{Huber}},
  \bibinfo{author}{\bibfnamefont{C.}~\bibnamefont{Mariani}}, \bibnamefont{and}
  \bibinfo{author}{\bibfnamefont{R.~B.} \bibnamefont{Vogelaar}}
  (\bibinfo{year}{2016}), \eprint{1602.04759}.

\bibitem[{\citenamefont{Vogel and Beacom}(1999)}]{Vogel:1999zy}
\bibinfo{author}{\bibfnamefont{P.}~\bibnamefont{Vogel}} \bibnamefont{and}
  \bibinfo{author}{\bibfnamefont{J.~F.} \bibnamefont{Beacom}},
  \bibinfo{journal}{Phys. Rev.} \textbf{\bibinfo{volume}{D60}},
  \bibinfo{pages}{053003} (\bibinfo{year}{1999}), \eprint{hep-ph/9903554}.

\bibitem[{\citenamefont{Huber}(2011)}]{Huber:2011wv}
\bibinfo{author}{\bibfnamefont{P.}~\bibnamefont{Huber}},
  \bibinfo{journal}{Phys. Rev.} \textbf{\bibinfo{volume}{C84}},
  \bibinfo{pages}{024617} (\bibinfo{year}{2011}), \eprint{1106.0687}.

\bibitem[{\citenamefont{Mueller et~al.}(2011)}]{Mueller:2011nm}
\bibinfo{author}{\bibfnamefont{T.~A.} \bibnamefont{Mueller}}
  \bibnamefont{et~al.}, \bibinfo{journal}{Phys. Rev. C}
  \textbf{\bibinfo{volume}{83}}, \bibinfo{pages}{054615}
  (\bibinfo{year}{2011}), \eprint{1101.2663}.

\bibitem[{\citenamefont{An et~al.}(2016{\natexlab{a}})}]{An:2015nua}
\bibinfo{author}{\bibfnamefont{F.~P.} \bibnamefont{An}} \bibnamefont{et~al.}
  (\bibinfo{collaboration}{Daya Bay}), \bibinfo{journal}{Phys. Rev. Lett.}
  \textbf{\bibinfo{volume}{116}}, \bibinfo{pages}{061801}
  (\bibinfo{year}{2016}{\natexlab{a}}), \eprint{1508.04233}.

\bibitem[{\citenamefont{Choi et~al.}(2016)}]{RENO:2015ksa}
\bibinfo{author}{\bibfnamefont{J.~H.} \bibnamefont{Choi}} \bibnamefont{et~al.}
  (\bibinfo{collaboration}{RENO}), \bibinfo{journal}{Phys. Rev. Lett.}
  \textbf{\bibinfo{volume}{116}}, \bibinfo{pages}{211801}
  (\bibinfo{year}{2016}), \eprint{1511.05849}.

\bibitem[{\citenamefont{Crespo-Anad{\'o}n}(2015)}]{Crespo-Anadon:2014dea}
\bibinfo{author}{\bibfnamefont{J.~I.} \bibnamefont{Crespo-Anad{\'o}n}}
  (\bibinfo{collaboration}{Double Chooz}), \bibinfo{journal}{Nucl. Part. Phys.
  Proc.} \textbf{\bibinfo{volume}{265-266}}, \bibinfo{pages}{99}
  (\bibinfo{year}{2015}), \eprint{1412.3698}.

\bibitem[{\citenamefont{Ashenfelter et~al.}(2016)}]{Ashenfelter:2015uxt}
\bibinfo{author}{\bibfnamefont{J.}~\bibnamefont{Ashenfelter}}
  \bibnamefont{et~al.} (\bibinfo{collaboration}{PROSPECT}),
  \bibinfo{journal}{J. Phys.} \textbf{\bibinfo{volume}{G43}},
  \bibinfo{pages}{113001} (\bibinfo{year}{2016}), \eprint{1512.02202}.

\bibitem[{\citenamefont{An et~al.}(2016{\natexlab{b}})}]{An:2016ses}
\bibinfo{author}{\bibfnamefont{F.~P.} \bibnamefont{An}} \bibnamefont{et~al.}
  (\bibinfo{collaboration}{Daya Bay}) (\bibinfo{year}{2016}{\natexlab{b}}),
  \eprint{1610.04802}.

\bibitem[{\citenamefont{Erickson et~al.}(2016)\citenamefont{Erickson,
  Bernstein, and Bowden}}]{Erickson:2016sdm}
\bibinfo{author}{\bibfnamefont{A.}~\bibnamefont{Erickson}},
  \bibinfo{author}{\bibfnamefont{A.}~\bibnamefont{Bernstein}},
  \bibnamefont{and} \bibinfo{author}{\bibfnamefont{N.}~\bibnamefont{Bowden}}
  (\bibinfo{year}{2016}), \eprint{1612.00540}.

\bibitem[{\citenamefont{{Carlson, J. and Bardsley, J. and Bragin, V. and Hill,
  J.}}(1999)}]{Carlson:1999}
\bibinfo{author}{\bibnamefont{{Carlson, J. and Bardsley, J. and Bragin, V. and
  Hill, J.}}}, in \emph{\bibinfo{booktitle}{{IAEA \protect{S}ymposium on
  \protect{I}nternational \protect{S}afeguards. Extended synopses}}}
  (\bibinfo{year}{1999}).

\bibitem[{\citenamefont{Abdurrahman et~al.}(1993)\citenamefont{Abdurrahman,
  Block, Harris, Slovacek, Lee, and Rodriguez-Vera}}]{abdurrahman1993spent}
\bibinfo{author}{\bibfnamefont{N.~M.} \bibnamefont{Abdurrahman}},
  \bibinfo{author}{\bibfnamefont{R.~C.} \bibnamefont{Block}},
  \bibinfo{author}{\bibfnamefont{D.~R.} \bibnamefont{Harris}},
  \bibinfo{author}{\bibfnamefont{R.~E.} \bibnamefont{Slovacek}},
  \bibinfo{author}{\bibfnamefont{Y.-D.} \bibnamefont{Lee}}, \bibnamefont{and}
  \bibinfo{author}{\bibfnamefont{F.}~\bibnamefont{Rodriguez-Vera}},
  \bibinfo{journal}{Nuclear science and engineering}
  \textbf{\bibinfo{volume}{115}}, \bibinfo{pages}{279} (\bibinfo{year}{1993}).

\bibitem[{\citenamefont{Hayes et~al.}(2015)\citenamefont{Hayes, Friar, Garvey,
  Ibeling, Jungman, Kawano, and Mills}}]{Hayes:2015yka}
\bibinfo{author}{\bibfnamefont{A.~C.} \bibnamefont{Hayes}},
  \bibinfo{author}{\bibfnamefont{J.~L.} \bibnamefont{Friar}},
  \bibinfo{author}{\bibfnamefont{G.~T.} \bibnamefont{Garvey}},
  \bibinfo{author}{\bibfnamefont{D.}~\bibnamefont{Ibeling}},
  \bibinfo{author}{\bibfnamefont{G.}~\bibnamefont{Jungman}},
  \bibinfo{author}{\bibfnamefont{T.}~\bibnamefont{Kawano}}, \bibnamefont{and}
  \bibinfo{author}{\bibfnamefont{R.~W.} \bibnamefont{Mills}},
  \bibinfo{journal}{Phys. Rev.} \textbf{\bibinfo{volume}{D92}},
  \bibinfo{pages}{033015} (\bibinfo{year}{2015}), \eprint{1506.00583}.

\bibitem[{\citenamefont{Huber}(2016)}]{Huber:2016fkt}
\bibinfo{author}{\bibfnamefont{P.}~\bibnamefont{Huber}},
  \bibinfo{journal}{Nucl. Phys.} \textbf{\bibinfo{volume}{B908}},
  \bibinfo{pages}{268} (\bibinfo{year}{2016}), \eprint{1602.01499}.

\bibitem[{\citenamefont{Hayes and Vogel}(2016)}]{Hayes:2016qnu}
\bibinfo{author}{\bibfnamefont{A.~C.} \bibnamefont{Hayes}} \bibnamefont{and}
  \bibinfo{author}{\bibfnamefont{P.}~\bibnamefont{Vogel}}
  (\bibinfo{year}{2016}), \eprint{1605.02047}.

\end{thebibliography}

\end{document}